# Experimental Measurement of Transverse Spin Dynamics in the Nonparaxial Focal Region

**Nitish Kumar,[1] Cyriac Raju,[2] Dinesh N. Naik[3] and Nirmal K. Viswanathan[1] ***

[1]School of Physics, University of Hyderabad, Hyderabad 500046, Telangana, India

[2]Department of Physics, Indian Institute of Science Education and Research Tirupati, Tirupati – 517619, Andhra Pradesh, India

[3]Department of Physics, Indian Institute of Space Science and Technology, Valiamala P.O., Thiruvananthapuram - 695 547, Kerala, India.

*Email: nirmalsp@uohyd.ac.in

*The superposition of complex optical fields in three-dimension (3D) is the basis of several non-trivial wave phenomena. Significant among them are the non-uniform polarization distribution and their topological character, leading to the emergence of transverse spin, transverse momentum, and their dynamics. These aspects are experimentally measured in the nonparaxial focal region of a circularly-polarized Gaussian input beam. A dielectric mirror, kept in the focal region, is axially scanned to obtain the phase and polarization variations in the retro-reflected output beam using an interferometer and spatially-resolved Stokes parameter measurements. The identification of phase and polarization singularities in the beam cross-section and their behaviour as a function of the mirror position enabled us to map and study the phase-polarization variations in the nonparaxial focal region. The lemon-monstar type polarization patterns surrounding the C-point singularity in the output beam are identified and tracked to study the transverse spin dynamics for right- and left- circular polarization of the input beam. Direct measurement of the input beam polarization helicity-independent and helicity-dependent aspects of the transverse and longitudinal spin dynamics in the nonparaxial focal region are the significant findings reported here. The proposed and demonstrated measurement method allows us to investigate the nonparaxial focal region in more detail and has the potential to unravel other intricate optical field effects.*

***Introduction.*** – The generic nonparaxial optical field is truly three-dimensional (3D). The realization of which has led to spatial structuring of optical beam-fields in a 2D plane and in 3D. These structured light beam-fields, in addition, reveal a close connection between the optical phase and polarization singularities, their topological characteristics, and are shown to carry linear and angular momenta, and optical forces [1 – 6]. All these features are significant both in the near-field and in propagating far-field, leading to realising 3D and 4D structured light beams and their applications [7 – 9]. Understanding the underlying connections stems from the polarization-dependent geometric phase, which is inherent to optical wave-field superposition, leading to the spin-orbit interaction (SOI) of light, and the topological aspects of vector wave-fields, that are to be represented on the Poincaré sphere and Majorana sphere [1, 10]. The experimental realization of polarization Möbius strip [11], a local increase in the degree of polarization due to tight focusing of unpolarized beam [12], and transverse spin and spin-momentum locking in polarized and unpolarized light [13, 14] are some of the significant findings in this context. Controlling and measuring these fundamental effects arising due to 3D polarization structures will enable diverse applications of nonparaxial optical fields.

As research thrust moves from the investigation of structured light beam-fields in the paraxial (2D) to the nonparaxial (3D) domain, theoretical-simulation activities anticipating several non-trivial spin-optical effects with hitherto unforeseen application potential are increasing rapidly [2, 4 – 6, 7]. The common physical origin of optical spin (helicity) -related topological effects in the evanescent, surface plasmon and propagating optical wave-fields [15, 16], allows one to explore such effects in the nonparaxial focal region. The optical field variations in the nonparaxial focal region are spatially nonuniform, and the polarization variations due to superposition of the field components and the topological structures in them are typically in the nano-scale. As a result, the experimental measurement of spin-optical effects is found lacking, largely due to significant challenges involved in measuring them. The most widely used experimental techniques in this context are limited to 3D raster scanning of a nano-particle kept in the nonparaxial focal region [17] and the scanning near-field optical microscope [18]. However, both these methods involve scanning of a nano-particle or a nano-tip along the $(x, y, z)$ axes and involve complicated field reconstruction process.

Reducing the 3D scanning to scanning only along the $z$ −axis with capability to simultaneously obtain the 2D field information in the $(x, y)$ plane provides significant measurement advantages. Axially scanning a plane mirror kept in the focal region of a lens has been used to obtain the 3D intensity [19], phase [20] and field components [21], but was declared unsuitable for tightly focused beams with strong longitudinal component [22]. Considering the recent realization that complex optical field characteristics vary continuously in 3D space and are threaded by phase and polarization singularities [23], we propose and demonstrate a new method to study the nonparaxial focal region of a high-numerical aperture (NA) lens. We make use of the presence of optical phase and polarization singularities in the focal region, due to superposition of the individual field components, to characterize and study the overall dynamics. The common origin of the EM dipole and optical beam fields [24], combined with the transversality requirement of the EM wave field, ensures that our measurement can be considered equivalent to the nano-particle and nano-tip scanning methods.

A uniformly polarized Gaussian laser beam, focused by a high-NA lens, is retroreflected by a dielectric multilayer-coated plane mirror kept in the focal region. The phase and polarization characteristics of the retroreflected output beam are measured using interferometric and Stokes polarimetry methods. Spatially-resolved phase and polarization variations in the output beam cross-section are measured using a CCD camera as a function of the axial scan of the mirror. This method allows us to obtain the complete 3D field distribution of the nonparaxial focal region. The lens-mirror-lens combination effectively transfers the near-field information available in the focal region to the far-field, ensuring 3D-to-2D transformation [25, 26]. We observe several significant phase and polarization features in the beam cross-section including phase singularities, nonuniform polarization distribution and C-point and L-line polarization singularities, all due to superposition of the optical field components in the focal region [27, 28]. These features are then used to study the input beam polarization helicity-dependent and helicity-independent longitudinal and transverse spin dynamics, and spin-momentum locking. The experimental measurements are supported by theoretical simulations, carried out based on the Richards-Wolf formalism [29] and the angular spectrum method [30], details of which are given in the Supporting Information section B.

***Theoretical aspects.*** – The coherence-polarization matrix of light is a 3×3 Hermitian matrix (eqn. 1), written in terms of the complex field amplitude ($E$) and relative phase ($\delta$) of the field components and contains all the measurable quantities, the intensity, and the state of polarization of the 3D EM wave-field [31]:

$$R = \langle \boldsymbol{E}(t) \otimes \boldsymbol{E}^\dagger(t) \rangle = \begin{bmatrix} \langle E_x^2(t) \rangle & \langle E_x(t)E_y(t)e^{-i\delta_{xy}(t)} \rangle & \langle E_x(t)E_z(t)e^{-i\delta_{xz}(t)} \rangle \\ \langle E_x(t)E_y(t)e^{i\delta_{xy}(t)} \rangle & \langle E_y^2(t) \rangle & \langle E_y(t)E_z(t)e^{i\delta_{yz}(t)} \rangle \\ \langle E_x(t)E_z(t)e^{i\delta_{xz}(t)} \rangle & \langle E_y(t)E_z(t)e^{-i\delta_{yz}(t)} \rangle & \langle E_z^2(t) \rangle \end{bmatrix} \quad (1)$$

Where, $\otimes$ is vector product, $\dagger$ is conjugate and transpose of the complex field $\boldsymbol{E}$ and $\langle ... \rangle$ is time average of optical fields. In the context of the work presented here, a uniformly polarized Gaussian beam of light focused by a high-NA lens consists of non-zero $(x, y, z)$ complex field components in the focal region [29, 30]. Details of the calculation of the diagonal and off-diagonal elements of the coherence-polarization matrix (eqn. (1)), and the simulated results are given in Section B of Supporting Information. The amplitude and the phase of the field components at the focal plane ($z = 0 \, mm$) for right circular polarized Gaussian input beam are the diagonal elements shown in Fig. B1. The complex field components also vary continuously in the focal region and at every transverse plane all the three components are present simultaneously [29, 30, 32, 33]. Thus, a coherent illumination of the high-NA lens leads to the presence of not only the diagonal elements (of eqn. (1)) but also the off-diagonal elements due to the superposition of all the three orthogonal field components. The superposition gives vectorial structure to the focal field, which has a strong dependence on the input beam polarization, the NA of the lens and the observation plane. From Fig. B1, one can clearly see that at the focal plane, the amplitude and phase of the field components have complicated spatial structure which can be measured via suitable experimental technique. It is important to point out here that the inability to directly measure the longitudinal field component in a propagating field can be circumvented by measuring its effects on the transversal components due to their superposition in the focal region. Measuring spatially-resolved scattered field sequentially at different planes in the focal region using a scanning nano-particle [17], or a nano-tip [18] or a lens [25] transfers the longitudinal component information to the transverse components, ensuring the 3D-to-2D transformation, and obtain the available focal field information. These methods have been used successfully to obtain the optical field components and their phase structure in the focal region of a high-NA lens. However, the methods have not been used to measure the field superposition effects in off-diagonal components of eqn. (1), which will complete the spin-orbit interaction effects [26, 32] in the nonparaxial focal region.

Here we propose to understand the nonuniform, polarization structured EM wave-field in the nonparaxial focal region, the resulting longitudinal and transverse spin components

($S_l$, $S_t$) and their dependence on the input Gaussian beam polarization helicity ($\sigma_i$) in terms of the off-diagonal components. Such fields are realized as due to the superposition of plane waves carrying helicity $\sigma_i$ for each local wavevector $\hat{k}_i$ and due to non-orthogonality of the plane-wave basis, the term $\sigma_{ij}\hat{k}_{ij}$ for the coupling of superposed plane waves $i$ and $j$ [15]. The longitudinal ($l$) and transverse ($t$) spin components are written as

$$S_l = \sum_i \hbar\sigma_i\hat{k}_i + \sum_{i\neq j} \hbar\sigma_{ij}\hat{k}_{ij} \qquad (2)$$

$$S_t = \frac{1}{2k^2}\nabla \times \prod \qquad (3)$$

In eqn. (2), the longitudinal component of spin $S_l$ is oriented parallel to the propagation direction $\hat{k}$, and the transverse spin $S_t$ is given by the vorticity of the kinetic momentum $\prod$ in eqn. (3). The total spin density is given by $S = S_l + S_t$ and its integral is calculated on the 2D transverse plane. More details and simulation results are given in Section B of Supporting Information. Including the coupling effects, the longitudinal spin is determined by the EM helicity ($\sigma$), whereas the transverse spin manifests in universal spin-momentum locking with its origin in the spatial inhomogeneity of the kinetic momentum ($\prod$). The transverse spin can be parallel to the mean wavevector and has both helicity-dependent and helicity-independent components [15, 16]. The evanescent wave geometry and nonparaxial focusing are the two scenarios where these non-trivial optical effects occur naturally and have been observed [13 – 15, 33, 34]. While many theoretical and simulation results are already available in these configurations, experimental observations of the effects are challenging and sophisticated that only a few groups pursue them.

*Experimental details.* – The experimental measurement of spin-optical effects arising due to the superposition of optical wave-fields in 3D, in the context of singular optics and optical angular momenta, is of significant current interest to unravel various novel optical effects and phenomena. In a beam of light with spatially varying phase and nonuniform state of polarization (SoP) this includes the presence of phase and polarization singularities [27, 28], spin-orbit interaction of light [35, 36], and effects related to transverse and longitudinal optical angular momenta [3 – 6]. Some of these optical phenomena are also proposed and observed in other forms of waves [37 – 40], highlighting the generality of these fundamental effects.

Schematic of the experimental setup is shown in Figure 1. The fundamental Gaussian beam from a stabilized He-Ne laser ($\lambda = 632.8\ nm$) is the monochromatic source used in the experiment. The laser beam is passed through a polarizer ($P_1$), a half-wave plate ($H_1$) and a quarter-wave plate ($Q_1$) to control the state of polarization (SoP) of the input light beam by suitably adjusting their orientation. After passing through a $50-50\ (R-T)$ non-polarizing cube beam-splitter ($BS_1$), the reflected part of the beam is focused using a $100\times$, 1.25 (oil) numerical-aperture (NA) and infinity-corrected microscope objective lens (OL, PLN, Olympus) of focal length $f = 1.60\ mm$ onto a plane mirror (DM). Since we use the lens in ambient conditions, the effective NA is taken to be < 1.0 ($\theta_m \sim 70°$) and for the Gaussian beam of diameter 1.45 mm entering the lens with an aperture diameter of 6.56 mm, the fill factor is calculated to be ~ 4.5 These high-NA lens parameters satisfy the Richards-Wolf formalism requirements [29].

The fully-reflecting dielectric multilayer-coated plane mirror is mounted on a motorized nano-positioner stage (MTS-25 Z8, Thorlabs) to enable sub-wavelength axial movement in the focal region of the lens. The input Gaussian beam diameter is measured, at the entrance aperture of the lens (OL), to be 1.45 mm, which is smaller than the 6.56 mm lens aperture diameter. This is ensured to avoid diffraction effects modifying the beam polarization characteristics during axial scanning, as reported by us in Ref. [41].

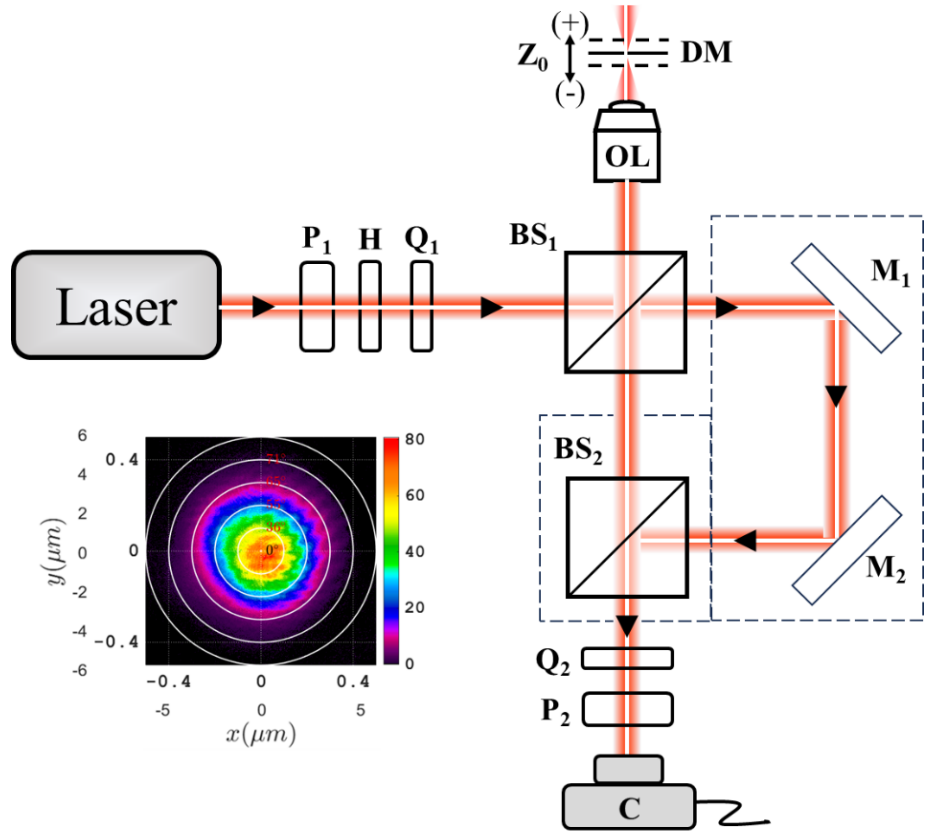

**Figure 1:** Schematic of the experimental setup used to measure the optical effects in the nonparaxial focal region. $P_{1,2}$, H and $Q_{1,2}$: polarizer, half-wave, and quarter-wave plates, $BS_{1,2}$: nonpolarizing 50-50 beam splitter, $M_{1,2}$: Mirror, C: CCD camera, OL: high-NA microscope objective lens, DM-Dielectric multilayer coated mirror. **Inset:** Reflected beam intensity for $Z \sim 0\ mm$. The two axes correspond respectively to the scale at the camera (in mm) and at the mirror (in μm) kept at the focal plane. The cone angles of the focused beam are indicated by white colour circles.

---

An independent reflectivity measurement of the mirror (DM) is carried out as a function of the angle of incidence and wavelength for $\pm 45^o$ polarization of the incident Gaussian beam. A weak polarization- and wavelength-dependent reflectivity is measured as a function of the angle of incidence in the range (0° − 70°) due to birefringence of the multilayer dielectric coating on the mirror [42]. The angle of incidence range of (0° − 70°) is larger than the angle subtended by the high-NA focused beam at the mirror of ~62°, corresponding to the lens focus. Replacing the mirror with a single surface silicon wafer high-reflector does not show birefringence effects in the reflected beam. The near-collimated retroreflected beam (after the lens) passes through the $BS_1$ and the intensity, phase and polarization characteristics of the transmitted beam are measured using a CCD camera (Thorlabs) kept in the far-field of the lens. The symmetry of the retroreflected beam intensity is established without polarization filtering. The measured beam intensity when the mirror is kept at ($z \sim 0\ mm$) focal plane of the lens (inset of Fig. 1) clearly shows the capability to resolve sub-wavelength features in the beam cross-section. The phase information of the reflected beam is obtained from the interferogram measured by constructing a two-beam interferometer (dotted line box in Fig. 1). Finally, using the ($Q_2 - P_2$) combination, kept in the output beam, oriented at ($0^o, 0^o$); ($0^o, 90^o$); ($0^o, +45^o$); ($0^o, -45^o$); ($90^o, +45^o$); ($90^o, -45^o$) allows the measurement of spatially-resolved Stokes parameters of the retroreflected beam [41]. From these measurements the phase and polarization singularity positions are identified [41] in the beam cross-section with sub-pixel resolution, for different axial positions of the plane mirror, kept within the Rayleigh range of the high-NA lens.

*Measurement of phase and polarization singularities.* – All the experimental and simulation results discussed correspond to either vertical- or right-circular polarized Gaussian input beam, unless mentioned otherwise. The spatially resolved measurements of the output beam intensity (Fig. 2 (a)), the corresponding interferogram and the Stokes parameters are carried out using the CCD camera kept at the Fourier plane of the lens. for three axial distances of $z = 0, \pm \lambda/2$ of the plane mirror kept in the high-NA lens focal region.

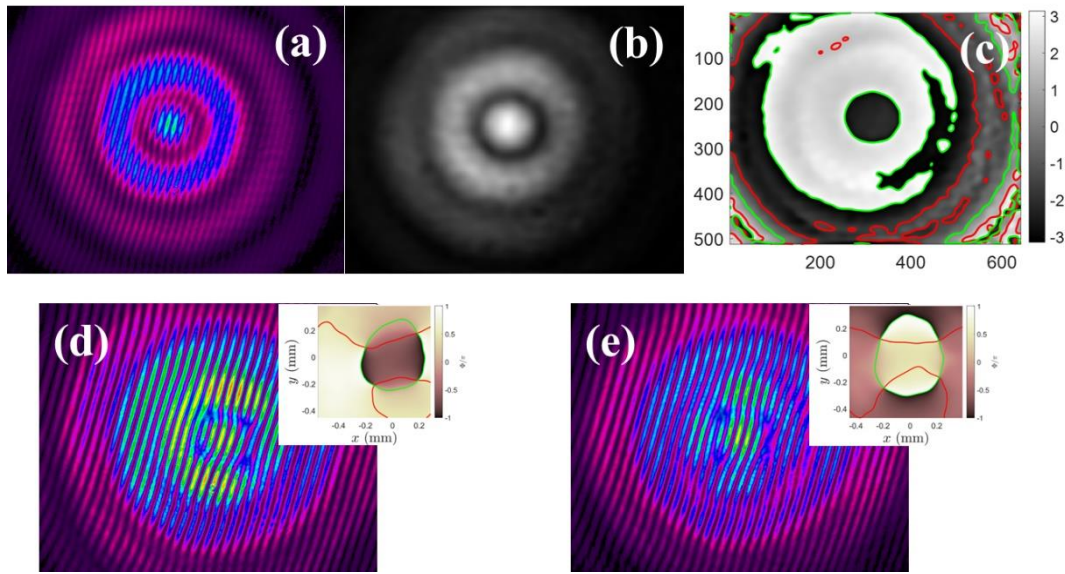

**Figure 2:** (a) Interferogram measured using the experimental setup in Fig. 1 for circular-polarized Gaussian input beam with the retroreflecting mirror kept at $z \sim 0\ mm$; (b), (c) The amplitude and phase extracted from the interferogram using Fourier transform (FT) method. (d), (e) Interferograms for vertical-polarized input Gaussian beam with the mirror at $z \sim \pm \frac{\lambda}{2} mm$, the insets show the phase extracted using the FT. Red and green colour lines in (c) and insets of (d) and (e) are the real and imaginary parts of the complex phase. The forklet patterns are clearly seen in the interferograms (d) and (e), corresponding to the positions where the real and imaginary parts of the complex phase intersect, as shown in the inset.

As the Rayleigh range of the high-NA lens used is $1.76\ mm$, an independent measurement is carried out first to identify the focal plane $(z = 0\ mm)$ location for positioning the mirror. For right-circular polarized Gaussian input beam, the retroreflected total beam intensity captured using the camera shows a bright central spot surrounded by weak, concentric dark and bright Airy rings ($Fig.\ 2a$). The Airy rings are regions of edge-type dislocation with $\pi -$ phase discontinuity [43, 44]. This is established by measuring the interferogram and performing the Fourier transform (FT) to get the amplitude and phase [45]. The interferogram measured with the mirror positioned at the focal plane, and the amplitude and phase extracted from it using FT are shown in (Fig. 2). Figure 2 (c) clearly shows circular edge-type phase dislocation. This position of the mirror, where the circular Airy ring pattern is measured, is marked as the focal $(z = 0\ mm)$ plane for all subsequent measurements. Next, for vertical polarized Gaussian input beam, as the mirror is moved away from the focal plane to $(z = \pm\lambda/2)$. The circular edge dislocation of the Airy ring gets deformed, leading to the appearance of quadrupole of optical vortices (OVs) of alternating sign [44]. The forklet pattern interferograms and the extracted phase maps are shown in Fig. 2 (d), (e) and their insets. The four OVs in the beam cross-section are rotated by $\pi/2$ for measurements before and after the focal plane as can be seen from their phase profiles. From the images, the $(x, y)$ position of the phase singular (vortex) points in the beam cross-section are calculated and given in Table 1, which will then be compared with the position where the C-point polarization singularities appear in the beam cross-section. The capability of our method to measure and identify edge- and vortex-type phase singularities in the focal region of the high-NA lens with sub-pixel accuracy is thus demonstrated. More detailed analysis of the phase singularity dynamics in the nonparaxial focal region and tracking of its trajectory are possible by scanning the mirror at $\lambda/10$ or less step size.

**Table 1:**
The spatial $(x, y)$ position of the phase singular optical vortex (OV) point and the C-point polarization singularity (PS) measured in the output beam cross-section, for the mirror positions $z_\pm = \pm\frac{\lambda}{2}\ \mu m$. The input beam is vertical polarized Gaussian beam. **Inset:** The singularities in the beam cross-section are numbered $(1 - 4)$. Red and green colour lines are the real and imaginary parts of the complex phase (Fig. 2 insets). The singularity positions (in $mm$) are calculated from the pixel numbers.

| [x, y] (mm) Number | OV (Z₋) | PS (Z₋) | OV (Z₊) | PS (Z₊) |
|---|---|---|---|---|
| 1 | (- 0.205, 0.074) | (- 0.155, 0.149) | (- 0.219, 0.121) | (- 0.181, 0.117) |
| 2 | (- 0.140, - 0.219) | (- 0.088, - 0.162) | (- 0.195, - 0.247) | (- 0.196, - 0.100) |
| 3 | (0.181, 0.177) | (0.149, 0.194) | (0.209, 0.107) | (0.164, 0.181) |
| 4 | (0.186, - 0.177) | (0.089, - 0.179) | (0.219, - 0.163) | (0.137, - 0.153) |

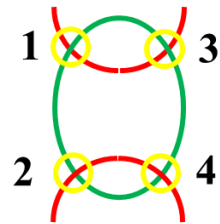

Next, we measure the state of polarization (SoP) of the retroreflected output beam to identify the C-point and L-line polarization singularities in the beam cross-section using CCD-based (spatially-resolved) Stokes polarimetry measurement [41, 46, 47]. For linear, vertical polarized Gaussian input beam, the six independent output beam intensities are measured using the CCD, for different fixed orientation of the quarter-waveplate – polarizer $(Q_2 - P_2)$ (Fig. A1). The measured Stokes images for the dielectric mirror kept at $z = 0\ mm$ are shown in Fig. A 1 (Supporting Information), measured for the $(Q_2 - P_2)$ combination of $(0^o, 0^o); (0^o, 90^o); (0^o, +45^o); (0^o, -45^o); (90^o, +45^o); (90^o, -45^o)$. Using these, the four Stokes parameters $(s_i, i = 0 - 3)$ are calculated (Fig. A 2) and using them the phase difference $(\Phi)$, polarization ellipticity $(\chi)$ and polarization ellipse orientation $(\psi)$ and variations in them in the beam cross-section are calculated (Fig. A3). It can be seen from these images that the SoP of the retroreflected output beam is nonuniform, supporting the fact that the optical field in the nonparaxial focal region of a high NA lens and retroreflected by a dielectric mirror is a superposition of the $(E_x, E_y, E_z)$ field components. These measurements and calculations are repeated for the three different axial positions $(z = 0, \pm \lambda/2)$ of the plane dielectric mirror kept in the focal region of the high-NA lens. Figure 3 (a) – (c) shows the spatially nonuniform SoP in the beam cross-section due to the superposition of all the field components present in the focussed-mirror-reflected beam [41, 47]. Further analysis of the data with the condition $(s_3 = 0, \pm 1)$ allows us to identify respectively the L-line and C-point polarization singularities in the beam cross-section and the streamlines connecting the polarization ellipses (Fig. 3) [28, 42, 47]. The quadrupole of C-point singularities is identified by white colour dots and red and green colour circles respectively for right and left circular polarization. Close to the beam center, the polarization ellipse field surrounding the C-points, as seen with the aid of the streamlines, are of the lemon-monstar type topological pattern [41, 47]. The transition region between the right and left elliptical SoP corresponds to L-line singularities. Figure 3 (a) shows that the nonuniform polarization structure observed near the first dark ring region with orthogonal C-points resembles a dipole, as seen from the streamlines. The C-points for $z = 0\ mm$ are close to annihilation, upon which the polarization singularities merge to become edge-type phase dislocation and the first Airy dark ring appears [43]. This corresponds to the condition where all the field components vanish and the SoP of the central bright spot becomes almost uniformly elliptical. On either side of this mirror position, the C-points move away from the center. The $(x, y)$ position of the C-point singularities that appear in the beam cross-section are given in Table 1, along with the phase singular points. As can be seen, the $(x, y)$ position of the phase and polarization singular points in the beam cross section for non-paraxially focused beams are different, as anticipated in Refs. [27, 28, 48]. It is important to note that the phase-singular (vortex) point and C-point polarization singularity merge with each other [48] only at the focal plane, where the Airy ring (intensity null), phase singularity and C-point singularity coincide. As can be seen, the $(x, y)$ position of the C-points close to the beam center come together, merge and move away from each other as the scanning mirror distance $(z)$ is varies from $-\lambda/2$ to 0 to $+\lambda/2$ indicating that the condition for the C-point singularity evolves in the focal region, indicating the 3D nature of the polarization field. It is important to note that the polarization singular C-points – L-lines are more closely trackable if one can vary the mirror at $\lambda/10$ or less step size, whereby the trajectory of the polarization singularities and the connecting 3D zeros of the EM fields can be tracked [23]. The capability for position-angle resolved measurement of the C-points in the beam cross-section at the sub-$\lambda$ scale is one of the significant advantages of our measurement scheme. In addition, as one goes radially outward from the beam axis, the lemon-monstar and the star type topological patterns surrounding the C-points alternate with each other and their dynamics in the nonparaxial focal region can be used for an in-depth study of their creation-annihilation studies more accurately than been attempted before. The variation of the polarization ellipse

orientation – ellipticity around the C-points also allows us to measure the *spin-Hall effect of light* (SHEL) at the nano-scale, a hitherto unforeseen advantage of our measurement scheme.

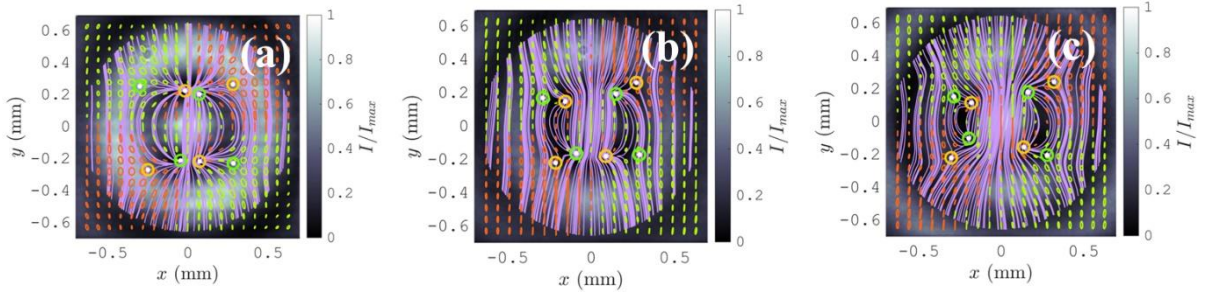

**Figure 3:** The state of polarization (SoP) of the focused-retroreflected beam, measured using the spatially-resolved Stokes parameters, are plotted on top of the beam intensity. The input beam is vertical-polarized Gaussian beam. The measurements are for 3 axial positions of the mirror: (a) $z = 0\ \mu m$, and (b), (c) $z = \pm\frac{\lambda}{2}\mu m$. Red and green colour respectively correspond to right- and left- circular / elliptical SoP. White colour dots with red / green colour circle indicate the C-point polarization singularity and the purple colour lines are the streamlines, connecting the tangent to the major axis of the polarization ellipses.

***Measurement of the role of polarization helicity on $(S_l, S_t)$.*** – One of the most significant recent findings in a 3D EM field with intrinsic polarization helicity ($\sigma$) is the emergence of transverse spin angular momentum (SAM) density and the spin-momentum (spin-current) coupling [2, 13-16, 30, 49-55]. Apart from the evanescent wave and surface plasmon wave systems, these effects have also been identified and measured in tightly focused circular polarized Gaussian and vector beams, by scanning a nanoparticle kept in the focal region [34, 50 – 55]. The presence of longitudinal and transverse spin angular momenta $(S_l, S_t)$ and their role in the nonparaxial beams´ behaviour are investigated here for right- and left- circular polarized Gaussian input beam.

The dynamic aspect of focused circular polarized optical beam-field in free-space consists of axial and azimuthal kinetic momentum (KM) components. The gradient of axial and azimuthal KM in the radial direction leads respectively to azimuthal and axial transverse spin [15]. Superposition of the axial transverse spin component with the longitudinal spin component constitutes the total SAM along the beam axis, which, when superposed with the azimuthal SAM results in Bloch-type spin Skyrmion [15]. The total axial SAM is helicity-dependent, and can be observed when the SoP of the incident Gaussian beam is changed from right- to left-circular polarization. The helicity-independent component of the transverse spin, arising due to spatial inhomogeneity of the KM, on the other hand results in spin-momentum locking, *i.e.,* KM-related vorticity. Here we study these 3D focal-field dynamics by tracking the C-point polarization singularity and the surrounding nonuniform polarization pattern in the output beam cross-section due to the superposition of the field components. We find that the polarization features in the beam cross-section evolve radially and azimuthally as a function of the propagation direction.

In the experimental setup ($Fig.$ 1), the SoP of the input Gaussian beam is now changed to right circular polarization to investigate helicity-related effects. This brings us to measuring spin-optical effects related to optical-field superposition in the 3D focal region. These are related to the SOI of light, leading to the measurement of helicity-dependent and helicity-independent longitudinal-transverse spin-momentum effects. These 3D aspects of the SAM in the focal region originate from the spatial-temporal inhomogeneity of optical spin and field momentum, leading to the appearance of spin-momentum, oriented orthogonal to the predominant direction of propagation of light and the rotation of field vectors in the longitudinal plane [49 – 55]. These unusual aspects of light's behaviour are also anticipated in a paraxial light field [16], observed in unpolarized light beam [13] and can be understood as due to the superposition of optical fields in the 3D.

In the circular polarization basis, the output beam in the same and orthogonal polarization measurement states correspond respectively to $l = 0$ Gaussian and $l = \pm 2$ Laguerre-Gaussian beams [32, 36]. The superposition of circular-polarized ($\sigma = \pm 1$) transverse field $(E_x \pm iE_y)$ components with the corresponding phase variations and the ($l = \pm 1$) phase vortex of the longitudinal field ($E_z$) component [Fig. B1] results in tilted (out-of-plane) polarization ellipsoid that evolves in the nonparaxial focal region. The spatial nonuniformity of the superposed field structure in the nonparaxial focal region results in spatial distribution of transverse components of the spin density which causes its rotation across the focal region [51 – 55]. Considering only the electric field contribution, the general definition for the optical spin density is $s_E \propto Im[E^* \times E]$ and the transverse components of the spinning electric field depend on their coupling with the longitudinal field component, given by $s_E^x \propto Im[E_y^* \times E_z]$ and $s_E^y \propto Im[E_z^* \times E_x]$ [52]. The transverse spin density for circular polarized input Gaussian beam is calculated and shown in the focal region [see Fig. B2. The required field coupling in experimental systems has been enforced either by using a gold nanoparticle [53] or a trapped mesoscopic particle [54, 55] or a plasmonic field probe [56]. In our experiments, the coupling of the 3D near-field components is achieved by angle-of-incidence dependent reflection at the dielectric multilayer coated mirror (DM). The angle of incidence varies from ($0^o - \sim 62^o$) due to the conical focus of the high-NA lens. Though the reflectivity due to multilayer dielectric coating on the mirror are in general assumed polarization-independent, our measurements confirm weak angle of incidence and wavelength-dependent birefringence and hence the mirror reflectivity. The weak field coupling effects in the focal region led to spatial variations and nonuniformity in the SoP in the output beam cross-section. The coupling effect also modifies the polarization variations in the output beam upon axial scanning of the mirror in the focal region. Thus, the coupling of optical near fields, in the nonparaxial focal region of the high-NA lens, enforced by the mirror reflection results in spatially nonuniform SoP in the beam cross-section.

The spatially nonuniform polarization measured in the focused-retroreflected output beam for right- and left- circular polarized Gaussian input beam are shown in Figure 4 (a) – (f). We identify the C-point and L-line polarization singularities satisfying the Stokes parameter condition $s_3 = \pm 1, 0$ in the nonuniformly polarized output beam cross-section. As compared to the four phase and polarization singular points in the beam cross-section for linear polarized input Gaussian beam, we measure only a pair of the singular points in the circular polarization basis [32, 36, 41]. Plotting the streamlines around the C-point polarization singularities allows us to identify lemon-monstar-star type topological pattern [41, 47, 48] for right- and left- circular polarized Gaussian input beam. The two C-point singularities in the beam cross-section, close to the beam axis, are of interest to us here. This is in contrast with the quadrupole of optical singularities in the beam cross-section, near the beam center, shown in Fig. 3 (a) –

(c), for linear vertical- polarized Gaussian input beam. The high-NA focusing of right- (left-) circular polarized input beam results in the appearance of ±2 charge vortex in the orthogonal left- (right-) circular polarization component due to spin-to-orbit conversion of light [32, 35, 36]. This in turn results in the appearance of the two C-point singularities in the output beam for circular polarized input beam. The near-axis C-point singularities and the lemon-monstar polarization structure in the output beam rotate counter-clockwise (CCW) as a function of the axial scan ($z$) of the mirror for right-circular polarized input beam. Along with the azimuthal rotation one also notices radial movement of the C-point singularities, moving close to the beam axis for $z = 0\ mm$ and away from the axis for $z = \pm\lambda/2$. These behaviours are understood as due to the coupling of the transverse SAM with the longitudinal OAM components and the resulting spin-orbit coupling in the focal region. The effects modifying the intensity and phase structure of the beam at the focal plane are also seen in the simulation results shown in Fig. B1. The measured rotation of the spatial distribution of the C-point singularity and the structure surrounding it are related to the underlying transverse spin density (Fig. B2). The behaviour has been explained as due to relative Gouy-phase shift between the transverse and longitudinal AM components [52]. In contrast, our earlier observation of the rotation of polarization structure around the C-point singularity is due to the Gouy phase difference between the orthogonal circular polarized Gaussian and Laguerre-Gaussian beams, a purely transverse effect [57]. The appearance and dynamics of the C-point singularity and the spatially nonuniform polarization ellipse structure measured in the beam cross-section relate to field coupling in the near-field, an effective 3D-to-2D transformation realized in the retroreflecting experimental configuration.

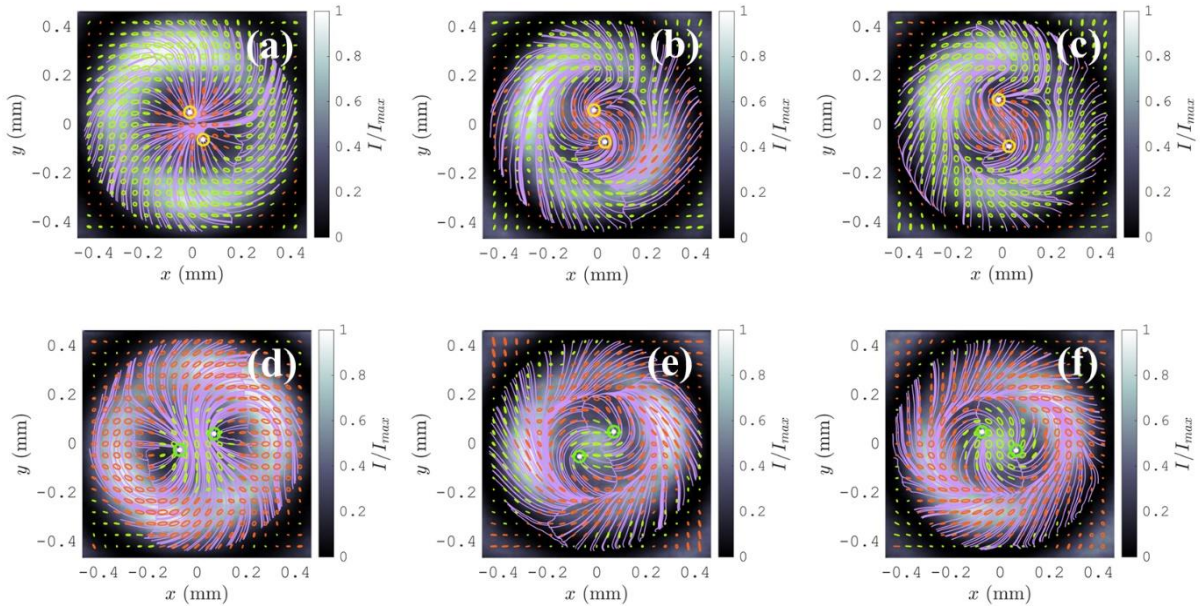

**Figure 4:** Nonuniform state of polarization (SoP) of the output beam, measured using the spatially-resolved Stokes parameters for (a) – (c) right- and (d) – (f) left- circular polarized Gaussian input beam, measured for 3 different axial position of the mirror. (a), (d) $z = 0\ mm$ and (b), (c) and (e), (f) $z = \pm\frac{\lambda}{2}\ mm$. Red and green colour respectively correspond to right- and left- circular / elliptical SoP. White colour dots with red / green colour circle indicate the C-point polarization singularity and the purple colour lines are the streamlines, connecting the tangent to the major axis of the polarization ellipses.
The radial and azimuthal movement of the C-point singularity and the surrounding streamlines are noticeable as a function of the mirror movement, indicative of the presence of transverse spin and its kinetic momentum.

Further assertion to the role played by the transverse spin density and its helicity dependence, resulting in the appearance and rotation of nonuniform polarization pattern with C-point singularity is established when the input beam polarization is changed from right- to left-circular. Figure 4 (d) – (f) shows the SoP in the beam cross-section, and its evolution as a function of the mirror position ($z = 0, \pm\frac{\lambda}{2} mm$) for the left-circular polarized input Gaussian beam. The streamlines enable us to visualize the flow pattern of optical fields. Changing the polarization helicity of the input beam from $\sigma = +1$ to $-1$ changes the longitudinal OAM component from $l = +1$ to $-1$ resulting in the appearance of helicity-independent lemon-monstar type polarization ellipse structure around the C-point singularities. The spatial location of the C-point singularities close to the beam center are different from that for right-circular polarization of the input beam (Table 1). The $z$−dependent clockwise (CW) rotation of the C-point singularity is opposite to that observed with right-circular input polarization and the nonuniform polarization around the C-points maintains the same lemon-monstar structure but in the orthogonal polarization. These measured characteristics and their behaviour confirm helicity-dependent SD and the spin-momentum coupling that appears in the high-NA lens´ focal region. The helicity-dependent and helicity-independent transverse spin behaviour clearly indicates the different role played by the three field components present simultaneously in the focal region and the effects arising due to their coupling. While additional details on the role played by the multilayer dielectric coating on the measured transverse spin effects are being worked out independently, we point out that using a single surface high-reflector (such as a silicon wafer) does not result in the required spatially nonuniform polarization in the focused-retroreflected output beam and hence all the other SD measurements reported here. We also would like to point out that the transverse spin density related effects can also be realized in transmission configuration, by using a transparent multilayer stratified structure instead of nanoparticle scattering systems reported [34, 52, 53].

***Summary.*** – Experimental measurement of polarization helicity-dependent and helicity-independent transverse spin dynamics is of significant fundamental importance. We proposed and demonstrated a retroreflection geometry to study the transverse spin dynamics in the nonparaxial focal region of a high-NA lens. This was realized by tracking the C-point singularities in the output beam as a function of the axial position of the dielectric mirror. The $100 \times$ optical magnification configuration allows us to map the nanoscale behaviour of the state of polarization in the nonparaxial focal region of the lens. The significance of the experimental scheme lends itself naturally to explore and understand the appearance, dynamics and topological aspects of complex vectorial field structures in the nonparaxial focal region, including but not limited to the Mobius strip, Skyrmions and other stable electromagnetic field structures.

***Acknowledgements.*** – The authors thank Science and Engineering Research Board (SERB), Government of India for financial support to this work and Upasana Baishya for critical reading of the manuscript. NK thanks Institution of Eminence (IOE), University of Hyderabad for additional fellowship support.

***Supporting Information.*** – Attached separately, provides all the necessary and additional experimental results, theoretical equations and the simulation results necessary to better understand the experimental results presented in the main text.

***References:***
[1] M.A. Alonso, Geometric descriptions for the polarization for nonparaxial optical fields: a tutorial. Adv. Opt. Photon., **15**, 176-235 (2023).
[2] Y. Shi, X. Xu, M.N. Vesperinas, Q. Song, A.Q. Liu, G. Cipparrone, Z. Su, B. Yao, Z. Wang, C.W. Qiu, and X. Cheng, Advances in light transverse momenta and optical lateral forces, Adv. Opt. Photon., **15**, 835 (2023).
[3] K.Y. Bliokh, and F. Nori, Transverse and longitudinal angular momenta of light, Phys. Rep., **592**, 1 (2015).
[4] K.Y. Bliokh et al., Roadmap on structured waves, J. Opt., **25**, 103001 (2023).
[5] Y. Shen et al., Roadmap on spatiotemporal light fields, J. Opt., **25**, 093001 (2023).
[6] H. Rubinsztein-Dunlop et al., Roadmap on structured light, J. Opt., **19**, 013001 (2017).
[7] E. Otte, Structured singular light fields, Springer thesis, Springer Nature Switzerland (2021).
[8] D. Marco and M.A. Alonso, Optical fields spanning the 4D space of nonparaxial polarization, arXiv:2212.01366v1.
[9] C.M. Spaegele, M. Tamagnone, S.W.D. Lim, M. Ossiander, M.L. Meretska, and F. Capasso, Topologically protected optical polarization singularities in four-dimensional space, Sci. Adv., **9**, 369 (2023).
[10] K.Y. Bliokh, M.A. Alonso and M.R. Dennis, Geometric phases in 2D and 3D polarized fields: geometrical, dynamical, and topological aspects, Rep. Prog. Phys. **82**, 122401 (2019).
[11] T. Bauer, P. Banzer, E. Karimi, S. Orlov, A. Rubano, L. Marrucci, E. Santamato, R.W. Boyd, and G. Leuchs, Observation of optical polarization Möbius strips, Science **347**, 964 (2015).
[12] K. Lindfors, et al. Local polarization of tightly focused unpolarized light. Nat. Photon. **1**, 228 (2007).
[13] J.S. Eismann, L.H. Nicholls, D.J. Roth, M.A. Alonso, P. Banzer, F.J. Rodríguez-Fortuño, A.V. Zayats, F. Nori and K.Y. Bliokh, Transverse spinning of unpolarized light, Nat Photon **15**, 156 (2021).
[14] M. Antognozzi, C. R. Bermingham, R. L. Harniman, S. Simpson, J. Senior, R. Hayward, H. Hoerber, M. R. Dennis, A. Y. Bekshaev, K. Y. Bliokh and F. Nori, Direct measurements of the extraordinary optical momentum and transverse spin-dependent force using a nano-cantilever, Nat. Phys., **12**, 731 (2016).
[15] P. Shi, L. Du, A. Yang, X. Yin, X. Lei and X. Yuan, Dynamical and topological properties of the spin angular momenta in general electromagnetic fields, Comms. Phys., **6**, 283 (2023).
[16] A.Y. Bekshaev, Transverse spin and the hidden vorticity of propagating light fields, J Opt. Soc. Am. A **39**, 1577 (2022).
[17] T. Bauer, S. Orlov, U. Peschel, P. Banzer, and G. Leuchs, Nanointerferometric amplitude and phase reconstruction of tightly focused vector beams, Nat. Photon., **8**, 23 (2014).
[18] N. Rotenberg and L. Kuipers, Mapping nanoscale light fields, Nat. Photon., **8**, 919 (2014).
[19] Y. Roichman, I. Cholis and D.G. Grier, Volumetric imaging of holographic optical traps, Opt. Expr., **14**, 10907 (2006).
[20] R. Juškaitis, Characterizing High Numerical Aperture Microscope Objective Lenses. In: Török, P., Kao, FJ. (eds) Optical Imaging and Microscopy, (2003). Springer Series in Optical Sciences, vol 87. Springer, Berlin, Heidelberg.

[21] I. Herrera and P.A. Quinto-Su, Measurement of structured tightly focused beams with classical interferometry, J. Opt. **25**, 035602 (2023).
[22] Y. Cai, Y. Liang, M. Lei, S. Yan, Z. Wang, X. Yu, M. Li, D. Dan, J. Qian, and B. Yao, Three-dimensional characterization of tightly focused fields for various polarization incident beams, Rev. Sci. Instr. **88**, 063106 (2017).
[23] A.J. Vernon, M.R. Dennis, and F.J. Rodriguez-Fortuno, 3D zeros in electromagnetic fields, Optica **10**, 1231 (2023).
[24] A. Aiello and M. Ornigotti, Near field of an oscillating electric dipole and cross-polarization of a collimated beam of light: Two sides of the same coin, Am. J Phys., **82**, 860 (2014).
[25] O.G. Rodríguez-Herrera, D. Lara, and C. Dainty, Far-field polarization-based sensitivity to sub-resolution displacements of a sub-resolution scatterer in tightly focused fields, Opt. Expr., **18**, 5609 (2010).
[26] O.G. Rodríguez-Herrera, D. Lara, K.Y. Bliokh, E.A. Ostrovskaya, and C. Dainty, Optical Nanoprobing via Spin-Orbit Interaction of Light, Phys. Rev. Lett. **104**, 253601 (2010).
[27] M.V. Berry, and M.R. Dennis, Phase singularities in isotropic random waves. Proc. R. Soc. Lond. A **456**, 2059 (2000)
[28] M.V. Berry, and M.R. Dennis, Polarization singularities in isotropic random vector waves. Proc. R. Soc. Lond. A **457**, 141–155 (2001).
[29] B. Richards, and E. Wolf, Electromagnetic diffraction in optical systems. II. Structure of the image field in an aplanatic system. Proc. R. Soc. Lond. A **253**, 358 (1959).
[30] L. Novotny, and B. Hecht, Principles of Nano-Optics (Cambridge University Press, 2012).
[31] J.J. Gil, Polarimetric characterization of light and media, Eur. Phys. J. Appl. Phys. **40**, 1 (2007).
[32] K.Y. Bliokh, E.A. Ostrovskaya, M.A. Alonso, O.G. Rodríguez-Herrera, D. Lara, C. Dainty, Spin-to-orbital angular momentum conversion in focusing, scattering, and imaging systems, Opt. Expr., **19**, 26132 (2011).
[33] A. Aiello, P. Banzer, M. Neugebauer, and G. Leuchs, From transverse angular momentum to photonic wheels, Nat. Photonics **9**, 789–795 (2015).
[34] M. Neugebauer, T. Bauer, A. Aiello, and P. Banzer, Measuring the transverse spin density of light, Phys. Rev. Lett. **114**, 063901 (2015).
[35] K.Y Bliokh, F.J. Rodríguez-Fortuño, F. Nori, and A.V. Zayats, Spin–orbit interactions of light. Nat. Photon. **9**, 796 (2015).
[36] K.Y. Bliokh, M.A. Alonso, E.A. Ostrovskaya, and A. Aiello, Angular momenta and spin–orbit interaction of nonparaxial light in free space. Phys. Rev. A **82**, 063825 (2010).
[37] D.A. Smirnova, F. Nori, and K.Y. Bliokh, Water-wave vortices and skyrmions, Phys. Rev. Lett., **132**, 054003 (2024).
[38] H. Zhang, Y. Sun, J. Huang, B. Wu, Z. Yang, K.Y. Bliokh, and Z. Ruan, Topologically crafted spatiotemporal vortices in acoustics, Nat. Commun., **14**, 6238 (2023).
[39] C Yang et al., Hybrid Spin and Anomalous Spin-Momentum Locking in Surface Elastic Waves, Phys. Rev. Lett., **131**, 136102 (2023).
[40] C. Shi et al., Observation of acoustic spin, Natl Sci. Rev. **6**, 707 (2019).
[41] N. Kumar, A. Debnath, and N.K. Viswanathan, Complex far fields and optical singularities due to propagation beyond tight focusing: combined effects of wavefront curvature and aperture diffraction, J. Opt. **26**, 045604 (2024).
[42] A. Yariv and P. Yeh, Electromagnetic propagation in periodic stratified media. II. Birefringence, phase matching, and x-ray lasers, J Opt. Soc. Am., **67**, 438 (1977).
[43] G. P. Karman, M. W. Beijersbergen, A. van Duijl, D. Bouwmeester, and J. P. Woerdman, Airy pattern reorganization and subwavelength structure in a focus, J. Opt. Soc. Am. A **15**, 884 (1998).

[44] J.N. Walford, K.A. Nugent, A. Roberts, and R.E. Scholten, High-resolution phase imaging of phase singularities in the focal region of a lens, Opt. Lett., **27**, 345 (2002).
[45] M. Takeda, H. Ina, and S. Kobayashi, Fourier-transform method of fringe-pattern analysis for computer-based topography and interferometry, J Opt. Soc. Am., **72**, 156 (1982).
[46] Goldstein D H 2011 Polarized Light, 3$^{rd}$ edn., (CRC Press, Taylor & Francis Group, LLC)
[47] V. Kumar, G. Philip and N. Viswanathan, Formation and Morphological Transformation of Polarization Singularities: Hunting the Monstar, J Opt., **15**, 044027 (2013).
[48] M.V. Berry, Index formulae for singular lines of Polarization, J. Opt. A: Pure Appl. Opt. **6**, 675–678 (2004).
[49] A. Y. Bekshaev, K. Y. Bliokh, and F. Nori, Transverse spin and momentum in two-wave interference, Phys. Rev. X **5**, 011039 (2015).
[50] S. Saha, N. Ghosh, and S. D. Gupta, Transverse spin and transverse momentum in structured optical fields, in Digital Encyclopedia of Applied Physics (Wiley, 2019), pp. 1–32.
[51] X. Pang and W. Miao, Spinning spin density vectors along the propagation direction, Opt. Lett., **43**, 4831 (2018).
[52] J. S. Eismann, P. Banzer, and M. Neugebauer, Spin-orbit coupling affecting the evolution of transverse spin, Phys. Rev. Res., **1**, 033143 (2019).
[53] M. Neugebauer, T. Bauer, A. Aiello, and P. Banzer, Measuring the transverse spin density of light, Phys. Rev. Lett. **114**, 063901 (2015).
[54] D. Pal, S. Dutta Gupta, N. Ghosh and A. Banerjee, Direct observation of the effects of spin dependent momentum of light in optical tweezers, APL Photonics **5**, 086106 (2020).
[55] R.N. Kumar, J.K. Nayak, S. Dutta Gupta, N. Ghosh and A. Banerjee, Probing Dual Asymmetric Transverse Spin Angular Momentum in Tightly Focused Vector Beams in Optical Tweezers, Laser & Photonics Reviews **18**, 2300189 (2024).
[56] K. G. Lee, H.W. Kihm, J. E. Kihm, W. J. Choi, H. Kim, C. Ropers, D. J. Park, Y. C. Yoon, S. B. Choi, D. H. Woo, J. Kim, B. Lee, Q. H. Park, C. Lienau, and D. S. Kim, Vector field microscopic imaging of light, Nat. Photonics **1**, 53 (2007).
[57] G.M. Philip, V. Kumar, G. Milione, N.K. Viswanathan, Manifestation of the Gouy phase in vector-vortex beams, Opt. Lett., **37**, 2667-2669 (2012).

<h1 style="text-align:center;">Supporting Information</h1>

# Experimental Measurement of Transverse Spin Dynamics in the Nonparaxial Focal Region

**Nitish Kumar,[1] Cyriac Raju,[2] Dinesh N. Naik[3] and Nirmal K. Viswanathan[1] ***

In the supporting information section, we present experimental measurement of spatially-resolved Stokes parameters and related calculations (in Section A) and in Section B the theoretical formulation and simulation results carried out to support the experimental results presented in the main text.

**Section A:**

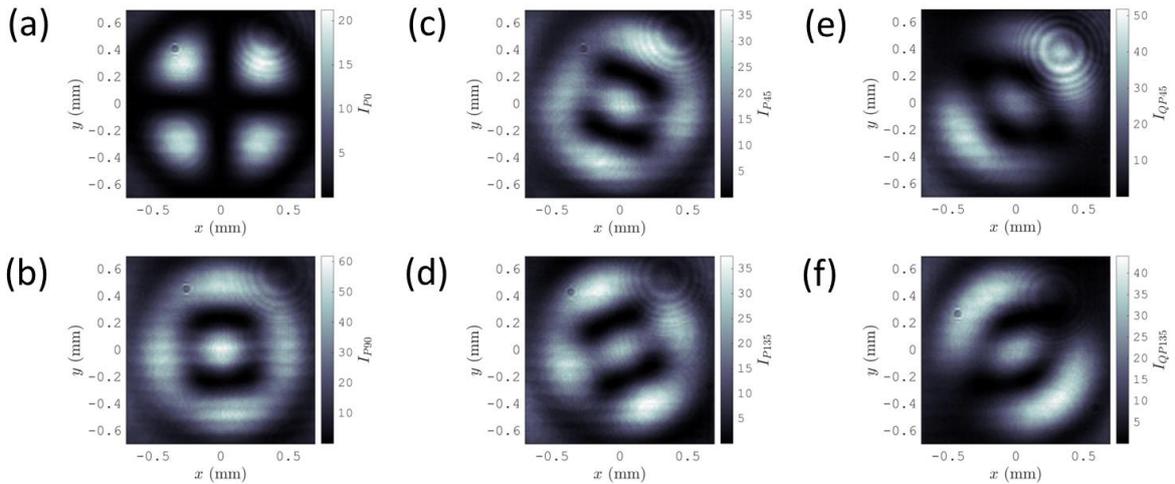

**Fig. A1:** Spatially-resolved intensity images for linear, vertically polarized input Gaussian beam measured after reflection at the mirror kept at $z \approx 0\ mm$ (Fig. 1). The six independent output beam intensity are recorded using the CCD camera for different fixed orientation of the quarter-waveplate – polarizer $(Q_2 - P_2)$ combination of (a) $(0^o, 0^o)$; (b) $(0^o, 90^o)$; (c) $(0^o, +45^o)$; (d) $(0^o, -45^o)$; (e) $(90^o, +45^o)$; (f) $(90^o, -45^o)$. Note here that $-45^o$ is the same as $135^o$ used for the analyser setting. It is clear from the images that the mirror-reflected Stokes images are quite complicated due to the interaction of the non-paraxially focused light beam with the multilayer dielectric coated mirror.

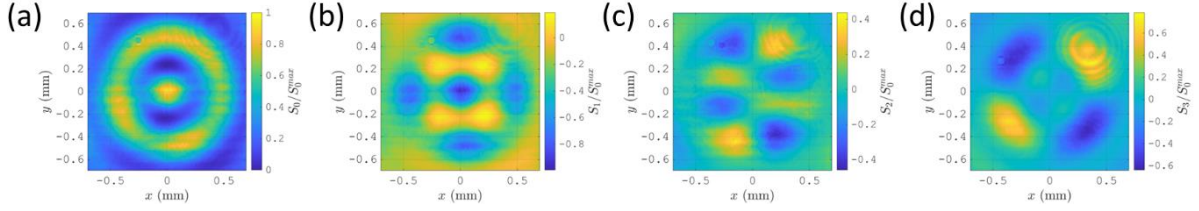

**Fig. A2:** Spatially-resolved Stokes parameters (a) – (d): $s_0 - s_3$, calculated using the images shown in Fig. A1 (a) – (f) in the formulae: $S_0 = I_0 + I_{90}$, $S_1 = I_0 - I_{90}$, $S_2 = I_{45} - I_{135}$ and $S_3 = I_{Q90P45} - I_{Q90P135}$. All the images are normalized with reference to $S_{0\,max}$; $s_0 = (S_0/S_{0\,max})$, $s_i = (S_i/S_{0\,max})$, $i = 1 - 3$. As can be seen, though theoretically one expects no $S_3$ component at the focal plane ($z = 0\,mm$), we clearly see all the Stokes parameters have spatial structure as the mirror is likely not kept exactly at the focal plane, (as mentioned in the main part of the manuscript) and the available field components at the focal plane are reflected by the multi-layer coated mirror which also exhibits birefringence upon reflection, depending on the cone angle of the incident beam.

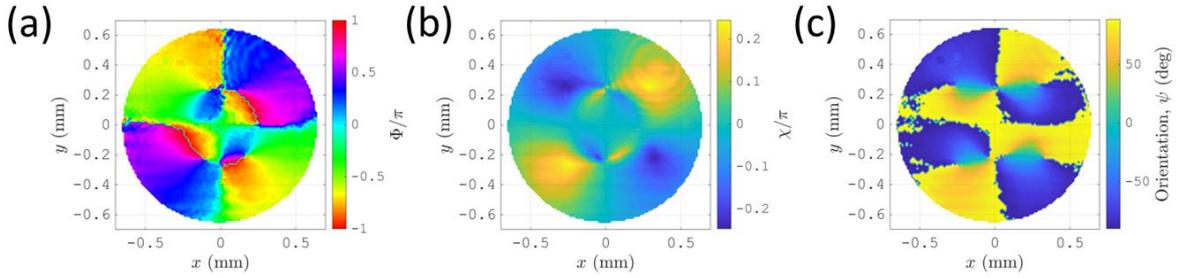

**Fig. A3:** Using the normalized Stokes parameters $(s_1, s_2, s_3)$ shown in Fig. A2 (b) – (d) we calculate spatially-resolved phase difference $\Phi = tan^{-1}\left(\frac{s_3}{s_2}\right), (-\pi \leq \Phi \leq \pi)$, ellipticity $\chi = \frac{1}{2}sin^{-1}\left(\frac{s_3}{s_0}\right), -\frac{\pi}{4} \leq \chi \leq \frac{\pi}{4}$ and ellipse orientation $\psi = \frac{1}{2}tan^{-1}\left(\frac{s_2}{s_1}\right), -\frac{\pi}{2} \leq \psi \leq \frac{\pi}{2}$ in the reflected beam cross-section. All these measured-calculated Stokes parameters are used together to create the spatially nonuniform polarization distribution in the reflected beam cross-section shown in Figs. 3 and 4 (main text) respectively for linear and circular polarized Gaussian input beams. The C-point and L-line in the beam cross-section are identified using $s_3 = \pm 1$ and $s_3 = 0$ conditions respectively.

## Section B:

The simulation of spin-orbit interaction (SOI) effects in the nonparaxial high-NA lens´ focal region, due to the focusing of Gaussian profile linear- and right- / left- circular polarized optical beam, is carried out using Python programming. Simulations are carried out to first plot and understand the transverse and longitudinal complex electric field components not only at the focal plane (Z = 0) but also in the focal region (Z = +/- λ). This allows us to calculate the diagonal and off-diagonal elements of the coherence (polarization) matrix (eqn. 1) in the nonparaxial focal region and understand the optical wave-field superposition effects. Subsequently, the complex fields are used to calculate the Stokes parameters and the SAM density in the nonparaxial focal region. All of this allows us to understand the experimentally measured phase-polarization singularities and their behaviour as the mirror is axially scanned in the focal region. These complex, intriguing, and non-trivial observations suggest very involved and intricate behaviour of optical fields due to the spin-orbit interaction in the 3D nonparaxial focal region.

**Methodology:** To simulate the nonparaxial focusing of a polarized Gaussian beam (GB) and to study the effects arising due to it, we developed a Python [1] code with the help of the PyFocus [2] and SymPy [3] modules. Here, we consider the GB to be propagating along the Z-axis, focused using a high NA lens. Considering the focal plane as the origin, we use modified PyFocus functions to calculate the amplitude and phase of the complex electric field components in the focal region. Richards-Wolf formalism [4] is used to calculate the complex electric field components in the focal plane for linear and circular polarized GB and simulate their behaviour in the nonparaxial focal region. The parameters used in the simulation are: the vacuum wavelength ($\lambda$) of the GB is 632.8 nm, the numerical aperture *(NA)* of the lens is 1.0, the refractive index (*n*) of the medium is taken to be 1.0, the aperture radius (*h*) is 0.5 mm, focal length *(f=nh/NA)* is 0.5 mm, incident field intensity ($I_0$) is 1.0 mW/cm$^2$, and the waist radius ($w_0$) of the input beam is set to 1.0 mm. The $\gamma$ and $\beta$ parameters respectively specify the input beam polarization ellipse orientation and the phase difference between the X and Y components of the electric field. These are set to 0º for linear X-polarization and 45° and 90°, respectively, for right- circular polarization. The total axial distance is set to 4.4 $\lambda$ (which includes Z = 0 mm focal plane) and is scanned with a step size of 0.02 $\lambda$. The XY plane is from -2 $\lambda$ to +2 $\lambda$ and is taken so to avoid any possible indexing errors due to ±2.2 $\lambda$ being the last plane in the integral matrices.

**Electric field components:** To simulate a focussed optical beam-field and study its characteristics, we first fix all the system parameters to the values given above. We also set the Z- and X- ranges and the step size, as they determine the three-dimensional region around the focal plane where the calculation is carried out for plotting. All the values in the plots are taken for each plane along the Z-axis (XY plane). For each plane along the Z- axis we find the semi-angle of the aperture $\alpha$, using $\alpha = sin^{-1}(NA/n)$. Then, we define the following functions:

$$g(\theta) = exp\left[-\left(\frac{\sin(\theta)f}{w_0}\right)^2\right] \quad\text{- (S1)}$$

$$f_1(\theta) = g(\theta)\sqrt{\cos\theta}\sin\theta\,[1+\cos\theta]J_0(k_r\sin\theta)exp[ik_z\cos\theta] \quad\text{- (S2)}$$

$$f_2(\theta) = g(\theta)\sqrt{\cos\theta}\sin^2\theta\,J_1(k_r\sin\theta)exp[ik_z\cos\theta] \quad\text{- (S3)}$$

$$f_3(\theta) = g(\theta)\sqrt{\cos\theta}\sin\theta\,[1-\cos\theta]J_2(k_r\sin\theta)exp[ik_z\cos\theta] \quad\text{- (S4)}$$

where, $f$ is the focal length, calculated using $f = nh/NA$, $J_v$ is the Bessel function of the first kind of $v^{th}$ order, $k_z = 2\pi z/\lambda$, where $z$ is the distance from the focus and $k_r = 2\pi r/\lambda$, where $r$ is the distance from the propagation axis.

The integral matrices are calculated for the Y = 0 plane in the X ≥ 0 region. For each point on the plane, say (X, 0, Z) where X ≥ 0 and Z ≥ 0, we calculate $k_z$ and $k_r$ values. These values are then used to integrate the $f_1$, $f_2$, and $f_3$ functions for $\theta$ from $0$ to $\alpha$. These are then respectively stored in as $I_1$, $I_2$, and $I_3$ matrices for each point (X, 0, Z) where X ≥ 0 and Z ≥ 0. Then, the complex conjugate of the +Z values are flipped and stored as the -Z values. The electric field components of the incident GB are calculated for the input parameters using $E_x = \frac{\sqrt{I_0}\cos(\gamma\pi f)}{\lambda}$, $E_y = \frac{\sqrt{I_0}\sin(\gamma\pi f)}{\lambda}$, $a_1 = E_x$, $a_2 = E_y e^{i\beta}$. To obtain the electric field components of the output beam for the XY planes at a particular Z = Z′ we first consider a point on the XY plane, say (X, Y, Z′) which is then converted to polar coordinates to obtain $r$ and $\phi$. The electric field components at (X, Y, Z′) are then calculated using

$$\boldsymbol{E_x}(z', y, x) = -ia_1[\boldsymbol{I_1}(z', r) + \boldsymbol{I_3}(z', r)\cos 2\phi] - ia_2\boldsymbol{I_3}(z', r)\sin 2\phi \qquad \text{- (S5)}$$

$$\boldsymbol{E_y}(z', y, x) = -ia_2[\boldsymbol{I_1}(z', r) - \boldsymbol{I_3}(z', r)\cos 2\phi] - ia_1\boldsymbol{I_3}(z', r)\sin 2\phi \qquad \text{- (S5)}$$

$$\boldsymbol{E_z}(z', y, x) = 2a_1\boldsymbol{I_2}(z', r)\cos\phi - 2a_2\boldsymbol{I_2}(z', r)\sin\phi \qquad \text{- (S7)}$$

This is repeated for each pair of *(x, y, z′)* and for each required XY plane, and the values are stored in the matrices $\boldsymbol{E_x}$, $\boldsymbol{E_y}$, and $\boldsymbol{E_z}$. The complex electric field components are then used to calculate all the other values of interest, such as the 3D coherence matrix elements, the Stokes parameters, and the SAM density, and obtain the plots.

**Coherence – Polarization matrix:** The complex electric field components are used to calculate the coherence (polarization) matrix elements (eqn. 1) shown in Fig. S1 for right-circular polarized input Gaussian beam. The diagonal elements, $|\boldsymbol{E_i}|^2$, give us intensities of the electric field components. These values are plotted on the required XY plane in terms of intensities (in kW/cm$^2$). The X- and Y- components exhibit comparable intensities and phase structures, except in their orientation. The intricate phase structures, containing multiple phase singularities distributed across the plane, correspond to dark spot in the respective intensity plot. The Z-component of the output beam is a vortex mode of helicity $l = +1$, a characteristic evident from its phase structure. The emergence of vortex structure in the output beam strongly suggests conversion from transverse SAM to longitudinal OAM, highlighting the presence of spin-orbit interactions within the total beam-field. Variations in the beam-field characteristics along the beam axis (Z-) is due to the component field variations and their superposition, both of which varies depending on the magnitude and phase of the field components. These effects are seen in the off-diagonal elements, which show coupling between the respective components in the focal region. This comprehensive analysis reveals the dynamic interplay between the spin and orbital angular momentum components in the optical field, underlining the intricate nature of spin-orbit interactions and their profound implications in modifying the resultant field and its behaviour.

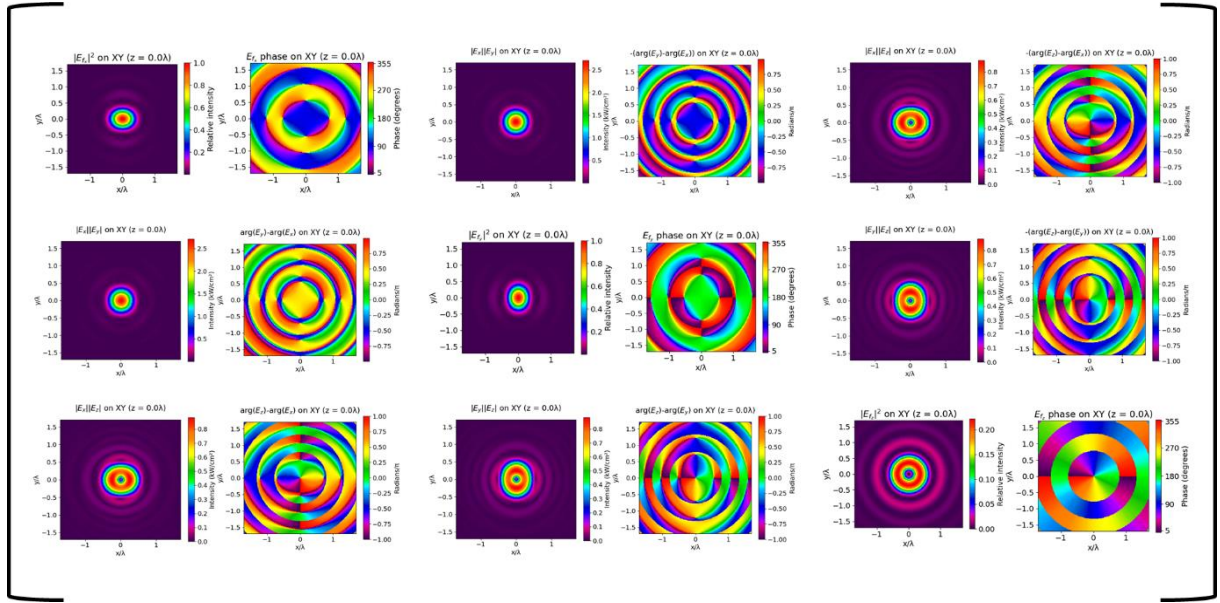

**Fig. B1:** The diagonal and off-diagonal elements of the coherence – polarization matrix (eqn. 1) and the corresponding phase maps calculated for right circular polarized input Gaussian beam at the focal plane (Z = 0).

**Spin AM density:** In addition to calculating the complex electric field components and the coherence (polarization) matrix elements due to superposition of the fields in the nonparaxial focal region, we delve deeper into the intricacies of light by calculating the spin angular momentum density. This provides critical insight into understanding the dynamics of light propagation, the resulting spin-orbit interaction of light and the appearance of transverse SAM in the nonparaxial focal region, all of which corroborate our experimental results.

The spin angular momentum (SAM) of light, associated with circular polarization state, represents a fundamental aspect of electromagnetic waves. Circularly polarized light has a helically varying electric field and possesses inherent SAM parallel to the direction of propagation of light. One can notice similarity between the SAM and the Stokes parameter $S_3$ as both represent and characterize elliptical polarization state of light. Understanding the SAM enriches our comprehension of light-matter interactions and fuels the development of advanced optical technologies with diverse applications across scientific disciplines. The physical quantity of relevance to this work here is the 3D aspect of the SAM, the SAM density, $\boldsymbol{J_S}$. The components of SAM density are calculated using [4, 5],

$$\boldsymbol{J_S} = \frac{\varepsilon_0}{4\omega_0} Im[\boldsymbol{E^*} \times \boldsymbol{E}] \qquad \text{- (S8)}$$

where, $\varepsilon_0$ is the permittivity in vacuum, $\boldsymbol{E^*}$ the complex conjugate of the electric field $\boldsymbol{E}$, and $\omega_0$ the angular frequency. As for the SAM density, we define $\boldsymbol{E}$ as a 3×1 matrix with components ($E_x$, $E_y$, $E_z$) and its complex conjugate $\boldsymbol{E^*}$. We then use eqn. (S8) to get the Cartesian components of the three orthogonal SAM density components using [4],

$$J_S = \begin{pmatrix} J_{Sx} \\ J_{Sy} \\ J_{Sz} \end{pmatrix}_{x,y,z} = \begin{pmatrix} Im[E_y^* E_z - E_y E_z^*] \\ Im[E_z^* E_x - E_z E_x^*] \\ Im[E_x^* E_y - E_x E_y^*] \end{pmatrix}_{x,y,z} \qquad - (S9)$$

The magnitude of the $J_S$ components is plotted on the required XY plane in terms of intensities (in kW/cm$^2$). Since we carry out here only a qualitative analysis, we omit the coefficients while calculating these values while maintaining the ratio of the values. And so, its units come out in terms of intensities. The resulting plots provide valuable insights into the distribution of SAM density in the XY plane and along the propagation axis.

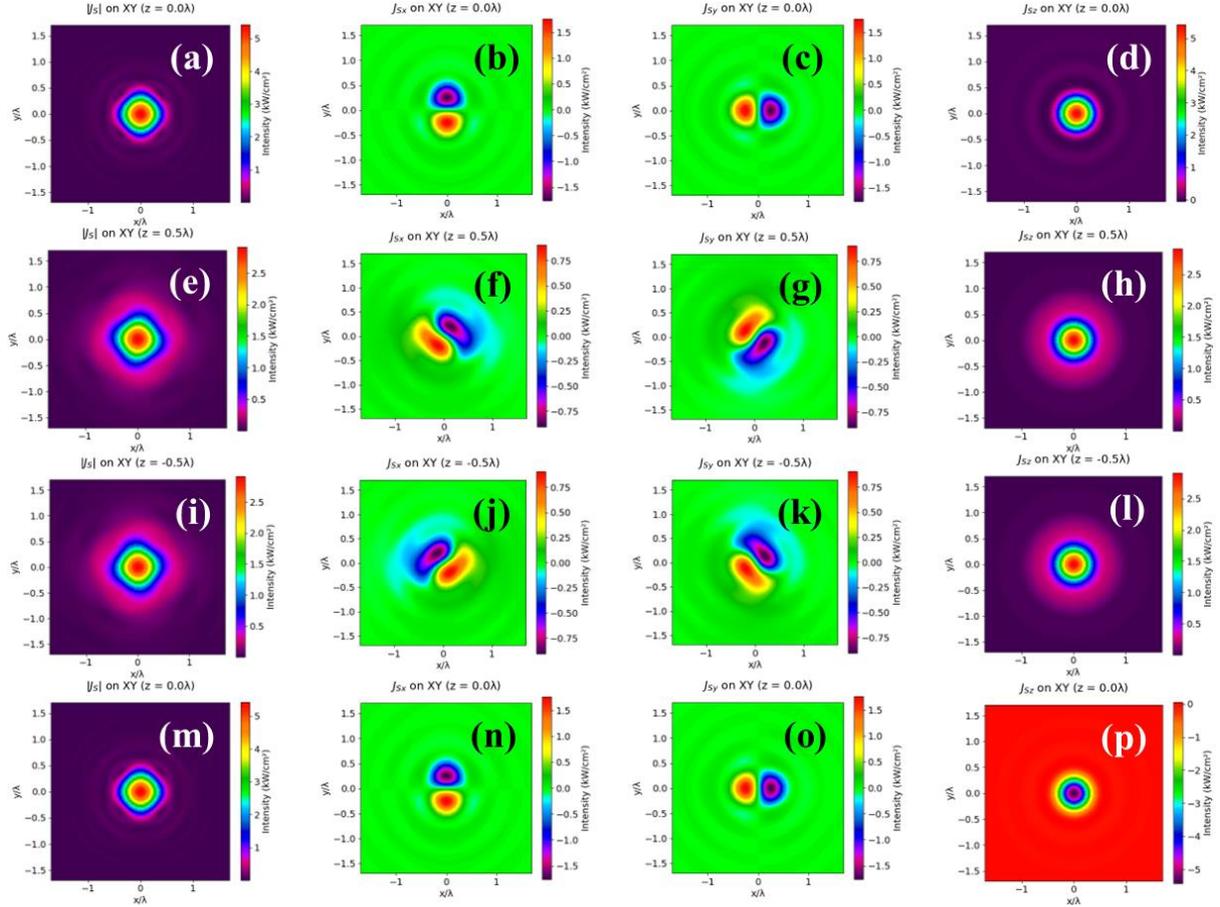

**Fig. B2:** The SAM density (a) $|J_S|$ and (b) – (d) components $(J_{Sx}, J_{Sy}, J_{Sz})$ for RCP input Gaussian beam in the focal plane, (e) – (h) and (i) – (l) at $\pm \lambda/2$ away, on either side of the focal plane. (m) – (p) is the same as (a) – (d) but for LCP input Gaussian beam.

**References:**

1. G. van Rossum, Python tutorial (CWI, Amsterdam, 1995).
2. Caprile F., Masullo L. A., and Stefani F. D., PyFocus – A Python package for vectorial calculations of focused optical fields under realistic conditions. Application to toroidal foci, Computer Physics Communications **275**, 108315 (2022).

3. Meurer A., Smith C. P., Paprocki M. et al., SymPy: symbolic computing in Python, PeerJ Computer Science **3**, e103 (2017).
4. Xiaohe Zhang et al., Understanding of transverse spin angular momentum in tightly focused linearly polarized vortex beams, Optics Express **30**, 4 (2022).
5. Aiello A., Banzer P., Neugebauer M. et al., From transverse angular momentum to photonic wheels, Nature Photonics **9**, 789–795 (2015).